\documentclass[conference]{IEEEtran}

\usepackage{cite}
\usepackage{amsmath,amssymb,amsfonts}
\usepackage{algorithm}
\usepackage{algorithmic}
\usepackage{graphicx}
\usepackage{textcomp}
\usepackage{xcolor}
\usepackage{comment}
\usepackage{float}
\usepackage{multirow}
\usepackage[hidelinks]{hyperref}
\renewcommand{\textcolor}[2]{#2}

\def\BibTeX{{\rm B\kern-.05em{\sc i\kern-.025em b}\kern-.08em
    T\kern-.1667em\lower.7ex\hbox{E}\kern-.125emX}}

\begin{document}

\title{On Evaluating Quantum Kernel Robustness for Low-Resource
Cross-Corpus Audio Deepfake Detection}
\author{
\IEEEauthorblockN{Lisan Al Amin, Lei Zhang, and Vandana P. Janeja}
\IEEEauthorblockA{University of Maryland, Baltimore County, Baltimore, Maryland, USA\\
salamin1@umbc.edu, leizhang@umbc.edu, vjaneja@umbc.edu}
}

\maketitle

\begin{abstract}
Detecting synthetic speech is critical for security, yet performance often
degrades when labeled data are scarce and evaluation conditions differ from
training. This study examines how quantum kernel methods compare with a
lightweight neural architecture for cross-corpus audio deepfake detection under
limited training data. We compare a Quantum Support Vector Machine (QSVM), a
classical support vector machine (SVM), and a multilayer perceptron (MLP), all
trained on frozen wav2vec 2.0 embeddings using a strict low-resource budget of
200 training samples. To match the qubit budget of near-term quantum hardware,
the embeddings are reduced to four dimensions with principal component
analysis, and the classical baselines use the same reduced features so that
differences between models reflect the decision function rather than the input
representation. Experiments on ASVspoof 2019, ASVspoof 5, the ADD 2023
Challenge, and the `In-the-Wild' dataset show that, under severe domain shift
from ASVspoof 2019 to ADD 2023, the MLP degrades to near-random performance,
with an area under the curve of approximately 50\% and an equal error rate of
50.0\%. In contrast, the QSVM maintains meaningful discrimination, achieving an
area under the curve of 76.0\% and an equal error rate of 27.0\%. This
advantage does not hold in every direction. When ADD 2023 is used as the
training source, the QSVM falls below chance on two of the three transfers and
 MLP becomes the stronger back-end. These findings suggest that quantum
kernel methods can provide a competitive alternative under severe cross-corpus
shifts and strict low-resource constraints, although they provide no consistent
advantage under near-domain transfer. We present these results as an empirical
characterization of the quantum kernel's inductive bias under distribution
shift rather than as a claim of quantum advantage, since the four-qubit kernel
studied here can be simulated exactly on classical hardware.
\end{abstract}


\section{Introduction}

Audio deepfakes created by modern text-to-speech and voice conversion systems pose a practical security risk, as synthetic speech can be used to spoof voice-based authentication and related verification pipelines. Community benchmarks have made this problem measurable at scale by releasing datasets, protocols, and standard metrics for countermeasure evaluation. ASVspoof 2019 introduced a large public database covering synthesized and converted speech together with established evaluation practice, including equal error rate and the tandem detection cost function \cite{todisco2019asvspoof,wang2020asvspoof2019db,todisco2018tdcf}. The ADD challenges further expanded the scope toward deep synthesis detection under diverse conditions~\cite{yi2022add2022,yi2023add2023}. More recently, ASVspoof 5 broadened the benchmark design with crowdsourced speech and explicit inclusion of deepfake and adversarial attack scenarios \cite{wang2026asvspoof5}. Across these evaluations, robustness remains a central concern because test audio can differ from training data in generation method, channel, and recording conditions.

Most recent countermeasures use deep neural networks trained on spectrograms or raw waveforms. RawNet2 is an example of an end-to-end raw waveform system designed for anti-spoofing \cite{tak2021rawnet2antispoof}, while AASIST models spoofing cues with integrated spectro-temporal graph attention \cite{jung2022aasist}. Transformer-based approaches have also been explored for audio classification, including audio spectrogram transformer-style encoders and related architectures \cite{gong2021ast,ulutas2023vitdeepfakeaudio}. However, state-of-the-art systems require massive datasets or expert curated datasets to converge \cite{aliras, isi}. Resource efficiency has become a critical research direction, with recent work investigating pruning techniques and the lottery ticket hypothesis to uncover minimal subnetworks required for deepfake detection \cite{amin2025uncovering}. We restrict training to 200 samples to simulate a ‘cold start’ scenario where a defender faces a novel attack. In this scenario, the defender of a voice authentication pipeline may hold only a handful of verified examples of a new synthesis or voice conversion method, with no time to curate a large corpus and no guarantee that the attack resembles the training distribution of an existing detector. Because voice authentication and call screening are interactive, a practical countermeasure must also return decisions within the live session, which favors lightweight back-ends that can be trained in seconds from a few hundred labels and score new audio at low latency. In parallel, self-supervised speech encoders provide transferable representations that can be reused as fixed front-ends. Wav2vec 2.0 is a widely adopted self-supervised framework that learns general-purpose speech representations from unlabeled audio \cite{baevski2020wav2vec2}. Given these strong representations, an open question is how much performance is determined by the front-end versus the decision function used in the back-end classifier, especially when labels are scarce.

Quantum machine learning provides an alternative way to increase classifier expressivity by changing the feature mapping rather than increasing the size of a neural network. Quantum kernel methods compute similarities after mapping inputs into a high-dimensional Hilbert space through a quantum feature map, enabling a support vector machine decision boundary in that induced space \cite{havlicek2019quantum,schuld2019feature}. Quantum kernel learning has been demonstrated on small-sample classification tasks in other domains \cite{wu2021quantumkernel}, and quantum support vector machines (QSVM) have also been studied for security-motivated audio analysis such as audio steganalysis \cite{venegas2024steganalysis} and robust audio deepfake detection under variable conditions \cite{amin2025reliable}. Practical experimentation is supported by hybrid quantum-classical software stacks such as PennyLane \cite{bergholm2020pennylane}. However, there is limited evidence on how quantum kernels compare to classical back-ends when evaluated on modern audio deepfake benchmarks under controlled protocols and strict label budgets. \textcolor{blue}{Our own prior study~\cite{amin2025reliable} is the closest to this one. It asked whether a quantum-kernel SVM stays reliable when acoustic conditions vary. Here we ask a different question. We fix the front-end and the input dimensionality so that only the decision function changes between back-ends, we cap training at 200 labeled samples, and we report every one of the nine cross-corpus transfers rather than the favorable ones. We also add calibration error and a fixed security operating point, and we read the outcome as an inductive bias of the feature map, not as a quantum advantage.}

In this work, we study the back-end question directly by fixing the representation and varying only the classifier. Utilizing frozen wav2vec 2.0 embeddings, we train three distinct models under identical data-scarce conditions: a QSVM, a classical SVM, and a lightweight Multilayer Perceptron (MLP)~\cite{baevski2020wav2vec2,cortes1995svm,bishop2006prml}. We evaluate on ASVspoof 2019, ADD 2023, ASVspoof 5, and `In-the-Wild' dataset \cite{todisco2019asvspoof,yi2023add2023,wang2026asvspoof5, muller2022does}. This study provides a controlled cross-corpus comparison of quantum, classical kernel, and neural back-end classifiers for low-resource audio deepfake detection. Our key contributions are:
\begin{itemize}
    \item \textbf{Cross-corpus low-resource robustness analysis:}
    Under a strict budget of $N=200$ labeled samples, we show that a lightweight MLP can degrade to near-chance performance under severe domain shift (e.g., ASVspoof 2019 $\to$ ADD 2023), while kernel-based methods retain stronger discrimination, with QSVM achieving AUC 76.04\% in this setting.

    \item \textbf{Competitive performance on modern benchmarks under strict supervision:}
    On ASVspoof 5, the proposed quantum-kernel pipeline achieves 8.00\% EER using only 200 labeled training samples, demonstrating that quantum kernels can remain effective in data-constrained spoofing detection.

    \item \textbf{Analysis of when the advantage holds and fails:} We report the full transfer matrix across all nine source and target pairs rather than favorable pairs only, showing that the QSVM advantage is confined to a subset of severe shifts, absent under near-domain transfer, and reverses when ADD 2023 serves as the training source, where the QSVM itself falls below chance on two of three targets. Because the four-qubit kernel studied here can be simulated exactly on classical hardware, we frame these findings as evidence regarding the inductive bias of the quantum feature map rather than a claim of quantum advantage.
    
\end{itemize}

The rest of this paper is organized as follows. Section~\ref{sec:method} details our method, Section~\ref{sec:setup} covers the experimental setup, Section~\ref{sec:results} presents the results and analysis, and Section~\ref{sec:conclusion} concludes the study.

\section{Evaluating Quantum Kernel: Our Method }
\label{sec:method}

We formulate deepfake detection as a binary classification task. Let $D = \{(x_i, y_i)\}_{i=1}^{N}$ denote a dataset where $x_i$ is the raw waveform and $y_i \in \{-1, +1\}$ is the class label. We focus on a low resource regime where the sample size $N$ is small. 

\begin{figure}[H]
  \centering
  \includegraphics[width=\columnwidth]{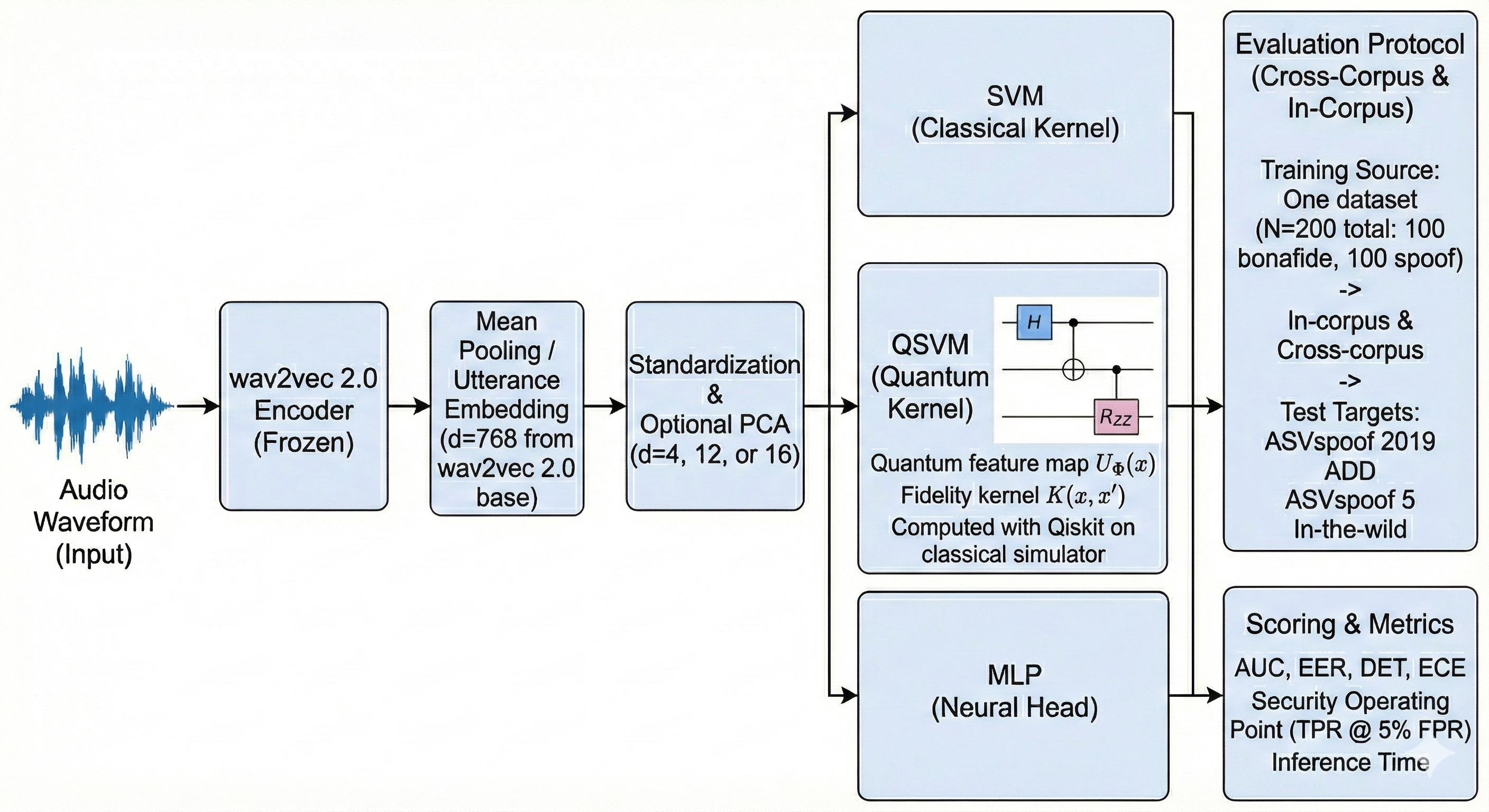}
  \vspace{-5mm} 
  \caption{Proposed Framework. Frozen wav2vec 2.0 features are reduced via PCA and passed to parallel back-ends.}
  \label{fig:methodology}
  \vspace{-2mm}
\end{figure}

\subsection{Feature extraction and dimensionality reduction}
We utilize a pre trained wav2vec 2.0 \cite{baevski2020wav2vec2} base model as a frozen front end, extracting representations specifically from the 12th and final hidden state. For a waveform $x$, the model produces hidden states which are temporally pooled to obtain an embedding $e \in \mathbb{R}^{D_{emb}}$. To map these features to the limited number of qubits available on Noisy Intermediate Scale Quantum (NISQ) devices~\cite{bharti2022noisy}, we apply Principal Component Analysis (PCA)~\cite{greenacre2022principal} to project $e$ onto a compact vector $z \in \mathbb{R}^{d}$, where $d$ corresponds to the number of qubits. The projection matrix is learned strictly on the training partition of $D$. Although projecting from 768 dimensions to 4 introduces an information bottleneck by retaining approximately 56.1\% of the total variance, this strict compression is necessary to satisfy current qubit constraints and rigorously tests the capability of the quantum feature map to learn a decision boundary from highly compressed representations. To avoid evaluation leakage, all data-dependent transforms, including feature standardization and PCA, are fit using only the training split in each fold, and then applied to the held-out split using the frozen parameters. For cross-corpus evaluation, the standardization and PCA are fit only on the source corpus training pool of $N=200$ samples, and the learned parameters are applied unchanged to every target corpus.

\subsection{Quantum kernel classification}
Instead of a classical kernel, we employ a quantum feature map to project data into a high dimensional Hilbert space. We utilize a ZZFeatureMap \cite{havlicek2019quantum,singh2025modeling}, which creates a state $|\Phi(z)\rangle$ using Hadamard gates for superposition and diagonal gates for data encoding. The quantum kernel entry is the fidelity between states:
\begin{equation}
\label{eq:quantum_kernel}
K_{q}(z_i, z_j) = \bigl\lvert \langle \Phi(z_i) \mid \Phi(z_j) \rangle \bigr\rvert^2.
\end{equation}
This kernel matrix is precomputed and passed to a classical SVM solver. 


\subsection{Classical baselines}

Our primary baselines are a classical SVM with a radial basis function (RBF)  kernel (tuned by grid search) and a lightweight MLP, both trained on the same reduced features $z$. We also evaluate additional classical models, such as Logistic Regression (LR), Random Forest (RF), Gradient Boosting (GB), and k-Nearest Neighbors (k-NN), in selected experiments to contextualize performance under the same low-resource setting. The MLP serves as a simple neural baseline for comparing kernel-based and neural decision functions under identical supervision.

\subsection{Design rationale and scope}
This study is designed as a controlled ablation of the decision function
rather than a search for the best possible deepfake detector. All back-ends
therefore receive exactly the same input: frozen wav2vec~2.0 embeddings
reduced by PCA to $d=4$ dimensions. This budget is imposed by the quantum
feature map on near-term hardware. The classical SVM and MLP are matched to
this input for parity. This is not because classical methods are restricted
to four dimensions. Baselines operating on the full 768-dimensional
embeddings would answer a different but equally useful question: the
performance of an unconstrained classical pipeline. But they would
reintroduce input dimensionality as a confound in any comparison of decision
functions. By fixing the representation and varying only the classifier,
observed differences can be attributed to how each decision function
organizes the same compressed evidence. Our results therefore position the
back-ends relative to one another within the compressed low-supervision
regime. They do not bound what full-dimensional classical systems can
achieve. We revisit this boundary when discussing limitations. Every other
factor in this design is also held fixed. Sample pools, fold splits, and
hyperparameter search ranges are identical across back-ends. The decision
function is the only variable under study.

\section{Experimental Setup}
\label{sec:setup}

We design experiments in which all methods operate on the same frozen wav2vec~2.0 embeddings and a fixed dimensionality reduction step. We enforce a strict low-resource setting with a labeled pool of $N=200$ samples per source corpus and report five-fold results under this fixed budget. The same sample pool and fold splits are used for QSVM, SVM, and MLP. This section describes the datasets, training protocol, and classifier configurations used throughout the study.


\subsection{Datasets and preprocessing}

We evaluate on four corpora covering standard benchmarks and real-world mismatch conditions: ASVspoof 2019 LA \cite{todisco2019asvspoof,wang2020asvspoof2019db}, ADD 2023 \cite{yi2023add2023}, ASVspoof 5 \cite{wang2026asvspoof5}, and the ``In-the-Wild'' dataset \cite{muller2022does}. All audio is resampled to 16~kHz to match wav2vec~2.0 input requirements \cite{baevski2020wav2vec2}, and the same preprocessing is applied across corpora to avoid introducing corpus-specific artifacts.

\subsection{Training protocol and classifier settings}

\textbf{QSVM:} Due to the high dimensionality of the wav2vec 2.0 embeddings ($d=768$), we first apply PCA to reduce the feature space to $d=4$ principal components. These 4 features are mapped to a 4-qubit quantum state using a ZZFeatureMap with circuit depth $L=1$ and full entanglement. We compute the quantum kernel matrix using the Qiskit framework and train a SVM with regularization parameter $C$ optimized via grid search using the precomputed kernel.
\textbf{Classical Baselines:} The classical SVM uses an RBF kernel with equivalent regularization ($C$ optimized via grid search) and automatic gamma scaling. The MLP consists of a single hidden layer with 128 units and ReLU activation, trained using Adam optimization with a learning rate of $10^{-3}$ for 50 epochs. For fairness, hyperparameters are selected on the training split only, using an inner cross-validation loop, and the same search ranges are used across back-ends.

For each training source corpus, we sample $N=200$ labeled training examples with equal class balance. For in-corpus evaluation, we report five runs with different random seeds and fixed fold construction. In cross-corpus experiments, models trained on this source pool are evaluated on the full target corpus evaluation split without any target adaptation. The training budget is fixed across QSVM, classical SVM, and MLP so that performance differences reflect the back end classifier rather than additional supervision.

\subsection{Evaluation metrics}
We report Accuracy, F1 score, Area Under the ROC Curve (AUC), and Equal Error Rate (EER). For security, we report the True Positive Rate (TPR) at a strict False Positive Rate (FPR) of 5\% \cite{yi2023add2023,delgado2021asvspoof2021evalplan}. We also report the Expected Calibration Error (ECE), computed by binning predictions by confidence and averaging the absolute difference between empirical accuracy and mean confidence in each bin~\cite{guo2017calibration}. We compute Accuracy and F1 by thresholding calibrated scores (Platt-scaled SVM probabilities) at 0.5, while EER is obtained by sweeping the decision threshold and selecting the operating point where false acceptance rate equals false rejection rate.

Quantum kernel computation is implemented using Qiskit framework on a statevector simulator, enabling exact and reproducible kernel evaluation. Input features are pre-scaled to $[0, \pi]$ and encoded into the Hilbert space via a ZZFeatureMap, which utilizes parameterized rotations and entangling gates to capture data correlations. The kernel entries are computed as the fidelity between statevectors corresponding to data pairs. This precomputed kernel matrix is passed to a classical Support Vector Classifier (SVC), where predictions are normalized to $[0,1]$ for consistent metric evaluation against neural baselines.

\section{Results and Analysis}
\label{sec:results}

We report mean and standard deviation over five folds under the fixed low-resource budget described in the experimental setup. Table~\ref{tab:in_corpus} summarizes the in-corpus performance.

\subsection{In corpus discrimination}
Under the strict budget, the QSVM achieves the highest mean discrimination on the majority of benchmarks (Table~\ref{tab:in_corpus}). To contextualize the role of nonlinearity of the quantum kernel, we performed an extended baseline analysis on the most recent benchmark, ASVspoof 5. We compared against linear (LR) and ensemble methods (RF and GB). While the linear baseline performed surprisingly well (EER 12.50\%), confirming that some spoofing cues are linearly separable, the QSVM reduced the error rate to 8.00\% (a 36\% relative reduction), suggesting that the QSVM can model nonlinear structure beyond the linear baseline under this protocol. \textcolor{blue}{The QSVM also shows the tightest spread over the five folds: its EER of 8.00 $\pm$ 2.45 compares with 12.50 $\pm$ 8.80 for the MLP, whose standard deviation is more than three times larger. We do not attach significance tests to these differences as the five folds are drawn from the same 200-sample pool and are not independent, so a paired test would overstate the evidence.}

\begin{table}[H]
    \centering
    \caption{\textbf{In-Corpus Performance (Low-Resource).} QSVM achieves the lowest EER (8.00\%) on the modern ASVspoof 5 benchmark.}
    \label{tab:in_corpus}
    \resizebox{\columnwidth}{!}{%
        \setlength{\tabcolsep}{2pt}
        \begin{tabular}{l l c c c c}
            \hline
            \textbf{Data} & \textbf{Model} & \textbf{Acc} $\uparrow$ & \textbf{F1 (\%)} $\uparrow$ & \textbf{AUC} $\uparrow$ & \textbf{EER} $\downarrow$ \\
            \hline
            \multirow{3}{*}{\textbf{ASV19}} 
             & MLP & 72.50 $\pm$ 4.47 & 70.40 $\pm$ 6.00 & 84.85 $\pm$ 2.73 & 24.00 $\pm$ 6.04 \\
             & SVM & 75.50 $\pm$ 5.10 & 74.40 $\pm$ 6.00 & 81.60 $\pm$ 3.70 & 24.50 $\pm$ 5.10 \\
             & \textbf{QSVM} & \textbf{78.50 $\pm$ 5.15} & \textbf{77.50 $\pm$ 5.00} & \textbf{84.90 $\pm$ 3.25} & \textbf{23.50 $\pm$ 6.82} \\
            \hline
            \multirow{7}{*}{\textbf{ASV5}} 
             & k-NN & 86.00 $\pm$ 4.36 & 88.43 $\pm$ 5.16 & 94.40 $\pm$ 3.72 & 11.50 $\pm$ 5.83 \\
             & RF & 87.00 $\pm$ 5.79 & 86.32 $\pm$ 6.50 & 94.00 $\pm$ 3.97 & 12.50 $\pm$ 5.48 \\
             & LR & 88.00 $\pm$ 4.58 & 87.69 $\pm$ 5.11 & \textbf{94.95 $\pm$ 2.87} & 12.50 $\pm$ 5.48 \\
             & GB & 88.00 $\pm$ 5.10 & 88.01 $\pm$ 5.38 & 94.65 $\pm$ 2.28 & 13.00 $\pm$ 7.81 \\
             & MLP & 86.00 $\pm$ 9.70 & 85.90 $\pm$ 9.00 & 94.00 $\pm$ 4.05 & 12.50 $\pm$ 8.80 \\
             & SVM & 87.00 $\pm$ 6.96 & 87.10 $\pm$ 7.00 & 94.35 $\pm$ 4.72 & 10.50 $\pm$ 7.97 \\
             & \textbf{QSVM} & \textbf{89.00 $\pm$ 4.64} & \textbf{88.40 $\pm$ 5.00} & 94.75 $\pm$ 2.95 & \textbf{8.00 $\pm$ 2.45} \\
            \hline
            \multirow{3}{*}{\textbf{ADD 2023}} 
             & MLP & 68.50 $\pm$ 13.4 & 70.80 $\pm$ 8.00 & 72.15 $\pm$ 19.0 & 28.00 $\pm$ 17.2 \\
             & SVM & 74.50 $\pm$ 7.14 & 75.00 $\pm$ 7.00 & 83.45 $\pm$ 5.50 & \textbf{23.00 $\pm$ 4.30} \\
             & \textbf{QSVM} & \textbf{75.00 $\pm$ 8.80} & \textbf{76.10 $\pm$ 8.00} & \textbf{84.25 $\pm$ 7.59} & 24.50 $\pm$ 11.5 \\
            \hline
        \end{tabular}%
    }
\end{table}

\subsection{Cross corpus robustness and deep learning degradation}
Our analysis reveals a critical vulnerability in deep learning baselines under severe domain shift. While the MLP baseline performs competitively on `near-domain' transfers such as ASVspoof 2019 $\to$ ASVspoof 5 (Table~\ref{tab:legacy_to_modern}), it degrades to near-chance performance on adversarial shifts. 
\begin{table}[!htb]
    \centering
    \caption{\textbf{Legacy $\to$ Modern Generalization.}
    Classical back-ends lead under this near-domain shift. The QSVM's advantage appears only under more severe shifts, discussed next.}
    \label{tab:legacy_to_modern}
    \resizebox{\columnwidth}{!}{%
        \setlength{\tabcolsep}{6pt} 
        \begin{tabular}{l c c c}
            \hline
            \hline
            \textbf{Model} & \textbf{EER (\%)} $\downarrow$ & \textbf{AUC (\%)} $\uparrow$ & \textbf{F1 Score} $\uparrow$ \\
            \hline
            MLP Baseline       & \textbf{16.00} & \textbf{89.05} & 0.746 \\
            Classical SVM (RBF)& 19.00 & 85.72 & \textbf{0.772} \\
            \textbf{QSVM} & 28.50 & 77.44 & 0.706 \\
            \hline
            \hline
        \end{tabular}%
    }
\end{table}
Specifically, when trained on ASVspoof 2019 and evaluated on the ADD 2023 Challenge, the MLP baseline essentially degrades, yielding an AUC and EER of 50.00\% (Table~\ref{tab:cross_corpus_master}). \textcolor{blue}{These are the values a random scorer would produce; on this transfer the MLP carries no usable signal.}

\begin{table}[t]
    \centering
    \caption{\textbf{Cross-Corpus Generalization.} Under severe domain shift (e.g., ASV19 $\to$ ADD 2023), MLP degrades (AUC $\approx$ 50\%) while QSVM retains discrimination (AUC 76\%).}
    \label{tab:cross_corpus_master}
    \setlength{\tabcolsep}{2pt} 
    \renewcommand{\arraystretch}{0.85} 
    \resizebox{\columnwidth}{!}{
    \begin{tabular}{l l | c c c | c c c | c c c}
        \hline
        \hline
        \multirow{2}{*}{\textbf{Train}} & \multirow{2}{*}{\textbf{Test}} & \multicolumn{3}{c|}{\textbf{MLP}} & \multicolumn{3}{c|}{\textbf{SVM}} & \multicolumn{3}{c}{\textbf{QSVM}} \\
        & & \textbf{EER} & \textbf{AUC} & \textbf{Acc} & \textbf{EER} & \textbf{AUC} & \textbf{Acc} & \textbf{EER} & \textbf{AUC} & \textbf{Acc} \\
        \hline
        \multirow{3}{*}{\textbf{ASV5}} 
         & In-Wild  & 39.00 & 66.93 & 63.50 & 33.00 & 69.81 & 67.00 & \textbf{31.00} & \textbf{74.64} & \textbf{69.00} \\
         & ASV19    & 24.00 & \textbf{86.19} & \textbf{73.00} & \textbf{22.00} & 84.58 & 72.50 & 31.50 & 79.79 & 68.50 \\
         & ADD 2023   & 50.00 & 50.00 & 50.00 & 44.00 & 57.85 & 50.00 & \textbf{40.00} & \textbf{62.36} & \textbf{55.50} \\
        \hline
        \multirow{3}{*}{\textbf{ASV19}} 
         & ADD 2023   & 50.00 & 50.00 & \textbf{50.00} & 57.00 & 37.39 & \textbf{50.00} & \textbf{27.00} & \textbf{76.04} & 49.00 \\
         & In-Wild  & 38.00 & 66.21 & 58.50 & \textbf{34.50} & \textbf{69.52} & \textbf{59.50} & 40.50 & 65.59 & 59.50 \\
         & ASV5     & \textbf{16.00} & \textbf{89.05} & 67.00 & 19.00 & 85.72 & \textbf{71.00} & 28.50 & 77.44 & 65.00 \\
        \hline
        \multirow{3}{*}{\textbf{ADD 2023}} 
         & ASV19    & \textbf{13.50} & \textbf{93.36} & \textbf{87.00} & 20.50 & 81.14 & 76.50 & 18.00 & 89.84 & 83.00 \\
         & ASV5     & \textbf{25.00} & \textbf{82.93} & 51.00 & 38.00 & 61.30 & \textbf{51.50} & 57.00 & 43.34 & 44.00 \\
         & In-Wild  & 47.00 & 50.89 & 47.00 & \textbf{43.50} & \textbf{52.95} & \textbf{49.50} & 52.50 & 45.18 & 45.50 \\
        \hline
        \hline
    \end{tabular}%
    }
\end{table}

\begin{figure}[H]
    \centering
    \begin{minipage}[b]{0.48\columnwidth}
        \centering
        \includegraphics[width=\linewidth]{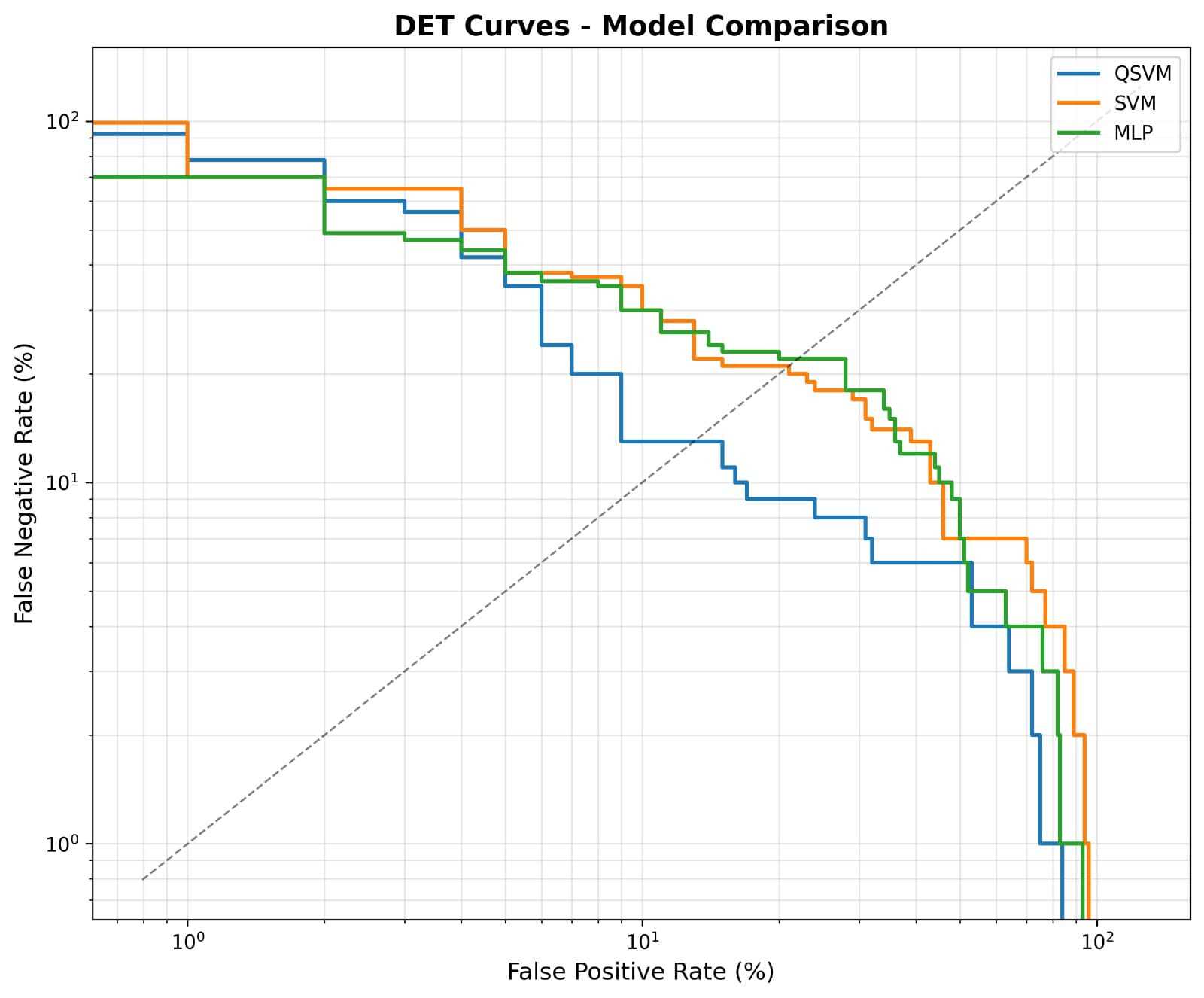}
        \centerline{\footnotesize (a) Adversarial Shift}
    \end{minipage}
    \hfill 
    \begin{minipage}[b]{0.48\columnwidth}
        \centering
        \includegraphics[width=\linewidth]{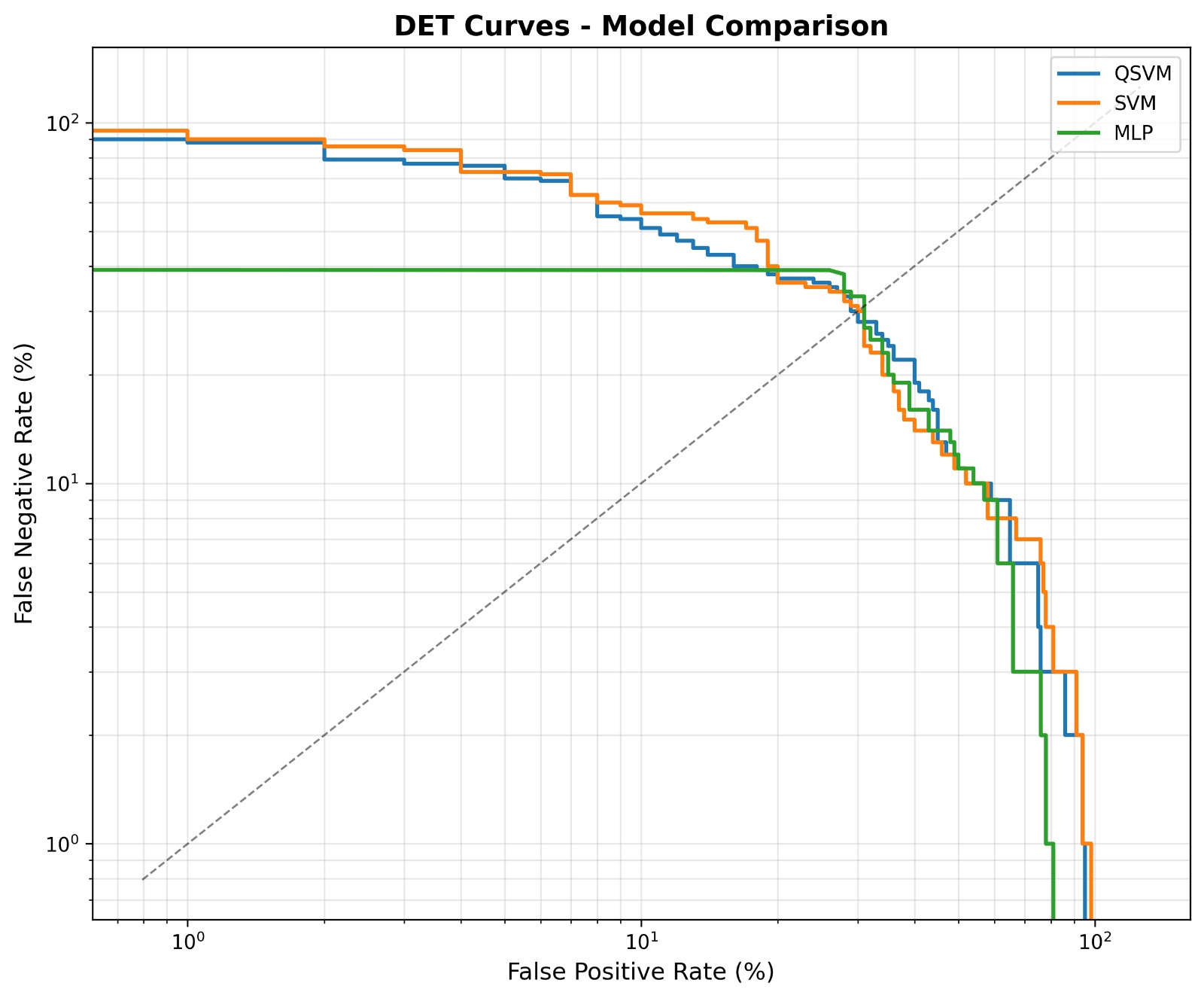}
        \centerline{\footnotesize (b) Real-World Noise}
    \end{minipage}
    
    \vspace{-3mm} 
    \caption{\textbf{Robustness.} QSVM (blue)  shows improved robustness on (a) adversarial shifts and (b) noisy data relative to the classical SVM (orange) and the MLP baseline (green) under the same protocol.}
    \label{fig:robustness_det}
\end{figure}

The classical SVM also performs poorly. Note that in this specific setting, the classical SVM yielded inverted decision scores (AUC $\approx$ 37\%). This may reflect a score inversion due to domain shift, resulting in an AUC worse than random and a corresponding EER of 57.00\%. The QSVM, however, retains meaningful discrimination, achieving an AUC of 76.04\% and an EER of 27.00\%. 
We observe a notable divergence between the strong AUC and reduced accuracy in this setting. This discrepancy arises because AUC is a threshold-independent metric measuring ranking quality, whereas accuracy relies on a fixed decision boundary established during training. Under severe domain shift, the calibration of decision scores drifts, causing the fixed threshold to fail even while the underlying separability (AUC) remains robust. This phenomenon is visually analyzed in Figure~\ref{fig:robustness_det}(a), where the MLP DET curve (green) follows the diagonal line of a random coin toss, while the QSVM curve (blue) bows significantly toward the origin.

\textcolor{blue}{The divergence between AUC and accuracy described above has a direct consequence for operational deployment, where a detector typically acts on a threshold fixed at training time rather than on a ranking. Table~\ref{tab:cross_corpus_master} shows what happens to that threshold under shift. From ASVspoof 2019 to ADD 2023, QSVM keeps an AUC of 76.04\%, yet its accuracy at the training-derived threshold is 49.00\%, no better than MLP. The scores still order bona fide and spoofed audio largely correctly, but the boundary between them has moved. The size of this loss can be read from the same table. The QSVM's EER of 27.00\% on this transfer is obtained by sweeping the threshold on the target data, so it shows what the same scores could deliver with a threshold set for that domain; the gap between that figure and 49.00\% accuracy is the cost of carrying the training threshold across the shift. A detector in this state can look healthy on ranking metrics while missing most spoofs at its deployed operating point, and in a voice authentication pipeline that means accepting synthetic speech as genuine. Two precautions follow for such a setting. The threshold should be re-estimated on a small labeled sample from the target domain rather than carried over from training. The sign of the scores should also be checked, since the classical SVM's AUC of 37.39\% on the same transfer shows that they can invert outright. Table~\ref{tab:security_analysis} gives the threshold-side view. At 5\% FPR the QSVM detects 15.0\% of spoofs against 4.98\% for the MLP, a real margin but a small one, and a reminder that most of the ranking advantage does not survive a strict fixed operating point.}

 \subsection{Reading the full transfer matrix}
 \label{sec:transfer-matrix}
The advantage above is real, but it is not the whole picture. Table
\ref{tab:cross_corpus_master} reports all nine source-to-target transfers, not
only the favorable ones. Across these nine pairs, the QSVM achieves the best
AUC on three. The MLP achieves the best AUC on four. The classical SVM leads
on the rest. The QSVM's advantage is concentrated on transfers where ADD 2023
is the target. In both directions of that shift, MLP or SVM AUC drops to
chance or below, while the QSVM retains meaningful discrimination.
The pattern does not hold in reverse. When ADD 2023 is the source rather than
the target, the QSVM itself degrades. ADD 2023 to ASVspoof 5 yields a QSVM
AUC of 43.34\%, below chance, with an EER of 57.00\%. ADD 2023 to In-the-Wild
yields a QSVM AUC of 45.18\%, also below chance. The MLP transfers well from
this same source, reaching an AUC of 93.36\% on ADD 2023 to ASVspoof 2019. A
plausible explanation is that the PCA and kernel fit on ADD 2023's
adversarially perturbed acoustics encode structure that does not generalize
back to cleaner corpora, and that the QSVM's rigid similarity structure, an
asset when the target is adversarial, becomes a liability in the reverse
direction.
We therefore state our claim narrowly. No single back-end dominates across
this transfer matrix. Kernel and neural decision functions fail in different
regions of it. The QSVM is the only back-end studied here that retains
discrimination when transferring into ADD 2023 from legacy data. This argues
for treating back-end choice as shift-dependent rather than fixed, and for
monitoring score orientation before deploying any back-end across a domain
change it has not been tested on.

\subsection{Why does the MLP degrade under severe domain shift?}

The large performance gap between the MLP and kernel-based methods under certain cross-corpus shifts suggests that the lightweight neural baseline is sensitive to distribution changes when supervision is limited.  Given the extreme scarcity of labeled samples, the MLP may underfit the target-relevant structure or overfit to dataset-specific regularities present in the training corpus. In this specific shift, the MLP yields near-chance discrimination (AUC $\approx$ 50\%), suggesting that under such extreme data scarcity, the learned neural decision rule does not transfer reliably to the target domain. While the MLP struggles to find a robust boundary under these constraints, the quantum kernel method preserves discriminative power, potentially related to the similarity structure induced by the kernel mapping.

Kernel methods (both classical and quantum) behave differently in this case. They rely on similarity structure induced by a fixed feature representation rather than learning a multi-layer decision function from scratch. This can make them less sensitive in some settings. Our results show that both quantum and classical kernels can be competitive. We therefore interpret the observed `degradation' primarily as a low-resource neural-baseline failure mode, and we view QSVM as a promising alternative whose benefits appear to be shift-dependent rather than universal.

\subsection{Inductive bias, not quantum advantage}
Two questions follow directly from our design. Is the quantum kernel
classically simulable? If so, why would it behave differently from an RBF
kernel at all?
The first answer is yes. The depth-one ZZFeatureMap on four qubits prepares a
state whose amplitudes are explicit functions of the input features. The
fidelity kernel in Equation~\ref{eq:quantum_kernel} sums over only $2^4 = 16$
basis states. This sum is computable exactly in time proportional to $2^n$ per
pair, where $n$ is the qubit count. This is exactly how we compute it here, on
a classical statevector simulator. No quantum-classical separation is claimed
or available at this width and depth. Rigorous kernel advantages known to
date require feature maps built around problems believed to be classically
hard \cite{liu2021rigorous}. Our contribution is an empirical account of an
inductive bias, not a complexity-theoretic claim.
The second question has a more tentative answer. The two kernels impose
different similarity geometries. With inputs scaled to $[0, \pi]$, the
quantum kernel is a bounded, oscillatory function of the input features. The
RBF kernel instead decays smoothly and monotonically with distance. Under
severe covariate shift, target points can move far from the training
manifold in Euclidean terms. An RBF kernel then assigns them near-zero
similarity to every support vector, and the decision function collapses
toward a constant. This may be why the classical SVM's scores invert under
the ASVspoof 2019 to ADD 2023 shift, and why the MLP's learned boundary fails
to transfer at all. The quantum kernel's oscillatory structure cannot collapse
in the same way. Shifted points still receive non-degenerate similarity
values to the support set. This is consistent with the QSVM's preserved AUC
even as its own calibration drifts.
We offer this as a hypothesis consistent with our measurements, not as a
proven mechanism. The reversal we report in
Section~\ref{sec:transfer-matrix} shows that the same rigidity can also
work against the QSVM when the source corpus, rather than the target, is the
adversarial one. A systematic account of when bounded similarity structure
helps and when it does not is a natural next step, and could be tested
directly by comparing the quantum and RBF kernels' alignment with the label
structure across each of the nine transfers in
Table~\ref{tab:cross_corpus_master}.

\subsection{Calibration and probability analysis}

We evaluate the reliability of model confidence using ECE, reported in Table~\ref{tab:calibration} for the modern ASVspoof 5 benchmark. We observe a distinct trade-off between discrimination performance and probability calibration. While the QSVM excels at ranking (highest AUC) and hard decision-making (lowest EER), its probability outputs are less calibrated than simpler baselines.  For instance, the k-NN and LR models achieve lower ECE scores (0.0670 and 0.1098, respectively) compared to the QSVM (0.1542).

\begin{table}[H] 
    \centering
    \caption{\textbf{Calibration (ASVspoof 5).} QSVM has best discrimination (lowest EER) but higher calibration error (ECE) than linear models.}
    \label{tab:calibration}
    \small 
    \setlength{\tabcolsep}{4pt} 
    \renewcommand{\arraystretch}{0.9} 
    \begin{tabular}{l c c}
        \hline
        \hline
        \textbf{Model} & \textbf{EER (\%)} $\downarrow$ & \textbf{ECE} $\downarrow$ \\
        \hline
        k-NN & 11.50 & \textbf{0.0670} \\
        LR & 12.50 & 0.1098 \\
        GB   & 13.00 & 0.1157 \\
        SVM     & 10.50 & 0.1252 \\
        RF       & 12.50 & 0.1691 \\
        \textbf{QSVM} & \textbf{8.00}  & 0.1542 \\
        \hline
        \hline
    \end{tabular}
    \vspace{-4mm} 
\end{table} 
Even the classical SVM (0.1252) is better calibrated than its quantum counterpart. This discrepancy suggests that standard Platt scaling, which fits a logistic sigmoid to decision scores, may be suboptimal for quantum kernel fidelities. The geometry of the quantum Hilbert space may produce concentrated fidelity values, which could skew the posterior probability mapping. While this does not affect the system's ability to correctly classify deepfakes (as evidenced by the superior EER of 8.00\%), it indicates that security analysts should interpret the \textit{raw confidence scores} of the quantum model with caution.

\subsection{Security analysis under bidirectional transfer}
Table~\ref{tab:security_analysis} evaluates models at a strict security operating point (TPR at 5\% FPR). We specifically isolate these three transfer pairs to evaluate performance under the most severe distribution shifts, namely bidirectional transfer between legacy and adversarial conditions, and generalization from modern benchmarks to uncontrolled real-world environments. The results demonstrate that the QSVM can maintain competitive security performance in some severe transfers across these extreme forward and reverse cross-corpus transfer scenarios, though this behavior is not uniform across all dataset pairs.

\begin{table}[!htb]
    \centering
    \caption{\textbf{Security Operating Point (TPR @ 5\% FPR).}
    QSVM is strongest on adversarial shifts (between ASV19 and ADD) under this operating point.}
    \label{tab:security_analysis}
    \resizebox{\columnwidth}{!}{%
        \setlength{\tabcolsep}{2.5pt} 
        \begin{tabular}{ll | c c c}
            \hline
            \hline
            \textbf{Train} & \textbf{Test} & \textbf{MLP} & \textbf{SVM} & \textbf{QSVM} \\
            \hline
            ASV19 & ADD 2023 & $4.98 \pm 0.02$ & $13.0 \pm 7.2$ & $\mathbf{15.0 \pm 4.5}$ \\
            ADD 2023 & ASV19 & $4.92 \pm 0.60$ & $15.0 \pm 1.6$ & $\mathbf{19.3 \pm 7.4}$ \\
            ASV5 & Wild & $8.75 \pm 2.45$ & $\mathbf{24.0 \pm 5.0}$ & $21.0 \pm 4.0$ \\
            \hline
            \hline
        \end{tabular}%
    }
\end{table}
When transferring from ASVspoof 2019 to ADD 2023, the QSVM achieves a TPR of \textcolor{blue}{15.0\%}, triple the performance of the MLP (4.98\%). Conversely, when transferring from ADD 2023 to ASVspoof 2019, the QSVM achieves a TPR of \textcolor{blue}{19.3\%}, numerically exceeding the MLP (4.92\%).

\subsection{Summary and Discussion}

Under a controlled setting with frozen wav2vec~2.0 embeddings and a strict low-resource training budget, kernel-based classifiers provide strong performance, and the QSVM is competitive across the evaluated benchmarks. A key observation is robustness under extreme cross-corpus shift. While the performance of the lightweight neural baseline (i.e., MLP) can degrade significantly, the quantum kernel retains meaningful separability and yields lower error rates in certain severe shifts. These results indicate that quantum kernels can be a practical alternative in low-supervision spoofing detection, particularly when generalization across datasets is the dominant requirement. At the same time, we note that the benefits of QSVM are case-dependent, which motivates further study of when quantum feature maps are most effective relative to classical kernels.

\subsection{Limitations}
The comparison here is bounded in a few ways. All back-ends see the same
four-dimensional PCA features, so we cannot say how a classical SVM or MLP
would perform on the full 768-dimensional embedding, only how each back-end
behaves under the same compression. Kernel entries are computed exactly on a
simulator, so shot noise and hardware runtime are not part of this study. The
QSVM's advantage is also not universal. It holds on transfers into ADD 2023
and disappears or reverses elsewhere. Training on ADD 2023 and testing on
ASVspoof 5 or In-the-Wild, the QSVM itself falls below chance while the MLP
transfers well. Its raw confidence scores are also less calibrated than
several classical baselines (Table~\ref{tab:calibration}) and should be read
with that in mind. \textcolor{blue}{This is also a comparison of back-ends behind one frozen front-end. We do not benchmark against fine-tuned state-of-the-art detectors such as Whisper- or AASIST-based systems~\cite{luo2024whisper}, and we do not report results on the recent large-scale multilingual corpora MLAAD~\cite{muller2024mlaad} and XMAD-Bench~\cite{ciobanu2026xmad}. All four corpora studied here are single-speaker; multi-speaker conversational deepfakes, for which a taxonomy and dataset now exist~\cite{ahmed2025multi}, are a further setting we do not cover. These would be needed before any claim about absolute performance, and all are left to future work.} Finally, all results use a single budget of $N=200$; we do
not know how these findings change at other budgets.

\section{Conclusion}
\label{sec:conclusion}

This work examined whether a quantum-kernel back end can improve audio deepfake detection in a strict low-resource setting with a fixed front end. Using frozen wav2vec~2.0 embeddings and an identical $N=200$ training budget, we compared three back-end classifiers: a QSVM, a classical RBF-SVM, and a lightweight MLP. 
Our results highlight a failure mode of the neural baseline under severe domain shift (ASVspoof 2019 $\to$ ADD 2023): the MLP degrades to random guessing (AUC $\approx$ 50.00\%, EER 50.00\%). In contrast, the QSVM retains meaningful discrimination in several low-resource cross-corpus settings, although its advantage is not consistent across all transfer directions. On ASVspoof 5, the QSVM achieves the best performance among the evaluated baselines under the same $N=200$ protocol (EER 8.00\%), compared with the MLP (12.50\%) and classical SVM (10.50\%). 
To address simulation scalability, we plan to implement quantum kernel approximations, such as Nystr\"{o}m methods, enabling training on larger datasets. We also aim to develop quantum-native calibration strategies to better align the model's high discriminative power with reliable probability estimates.

\section*{Acknowledgment}
\textcolor{blue}{This work is funded by the National Science Foundation Award \#2346473 ``CIRC: DEV: Community Infrastructure for Advancing Audio Deepfake Detection''.}

\bibliographystyle{IEEEtran}
\bibliography{refs}

\end{document}